\documentstyle[aps,amssymb]{revtex}

\begin{document}

\title{The attractors in sequence processing neural networks\thanks{%
Published in Int. J. mod. Phys. C11, 33(2000)}}
\author{Yong Chen, Ying Hai Wang and Kong Qing Yang \\
\textit{Physics Department of Lanzhou University, China (730000)}\\
\textit{E-mail: ychen@lzu.edu.cn}}
\maketitle

\begin{abstract}
The average length and average relaxation time of attractors in sequence
processing neural networks are investigated. The simulation results show
that a critical point of $\alpha $, the loading ratio, is found. Below the
turning point, the average length is equal to the number of stored patterns;
conversely, the ratio of length and numbers of stored patterns, grow with an
exponential dependence $\exp \left( A\alpha \right) $. Moreover, we find
that the logarithm of average relaxation time is only linearly associated
with $\alpha $ and the turning point of coupling degree is located for
examining robustness of networks.

\end{abstract}

\section{\protect\vspace{0in}{\protect\large Introduction}}

The dynamic behavior of Hopfield model$^1$, a global symmetric coupling
neural network, is relatively simple: the system relaxes to a fixed point
corresponding to a stable patterns$^2$. The latest result of its maximal
stored capacity $\alpha _s$ given by Volk$^3$ is $0.143\pm 0.002.$ Because
the synaptic connections in real biological neural networks have a highly
degree of asymmetry and a real human idea consists of a set of patterns,
many modification of Hopfield model have been designed$^{4-6}$. In general,
an asymmetric connection strength may arise to recall a series of patterns
or even chaos behavior, and cannot be studied by conventional statistical
methods because it disobeys detailed balance.

In this paper, we study the behavior of attractors in a sequence processing
model$^5$, a fully connected Ising spin model, through numerical simulation
with deterministic parallel dynamics. Moreover, the robustness is examined
by locating critical coupling degrees since the loss of synaptic connections
in the human brain may occur because of brain damage$^7$.

\section{\protect\large Model definition}

We consider a sequence processing neural network model described by the
following equations$^{4-6}$ 
\begin{equation}
s_i\left( t+1\right) =\mathrm{sgn}\left( \sum\limits_jJ_{ij}s_j\left(
t\right) \right)
\end{equation}
where Ising spin $s_i\left( t\right) \in \left\{ -1,1\right\} $ represents
the firing state or resting state and $\mathrm{sgn}\left( x\right) $ is
updating function corresponding with signum function which if $x>0$, we get $%
1$; if $x<0$, we get $-1$. The neurons evolve their states simultaneously
with deterministic parallel dynamics and the time step is $1$ in all our
work.

The synaptic connection matrix $\mathbf{J}$ can be given by$^{4,8}$ 
\begin{equation}
J_{ij}=\frac 1N\sum_{\mu =1}^qc_{ij}\xi _i^{\mu +1}\xi _j^\mu
\end{equation}
for a diluted model consisting of $N$ neurons. The $q$ vectors $\xi ^\mu
=\left\{ \xi _1^\mu ,\xi _2^\mu ,\ldots ,\xi _N^\mu \right\} \in \left\{
-1,1\right\} ^N$ are randomly and independently stored patterns in neural
networks. It means the stored patterns are organized in one $q$-period
cycle. It can be introduced by definition $\xi _i^{q+1}=\xi _i^1$ in $(2)$.
The dilution factors $c_{ij}=0$ or 1 are independent identically distributed
random variables. They are selected by 
\begin{equation}
\mathrm{if}\quad z\leq d,\ \mathrm{then}\quad c_{ij}=1;\quad \mathrm{else}\
c_{ij}=0
\end{equation}
where $z\in [0,1]$ is a random number and $d\in [0,1]$ is the coupling
degree of networks. For $c_{ij}=1$ or $d=1$, the model restore to a standard
(Hopfield-Hebb) sequence processing model.

Comparing with standard Hopfield model, the synapses of equation $\left(
2\right) $ are insufficient for the generation of stable fixed point in
phase space, since they combine pattern $\mu$ with pattern $\mu+1$.
Nevertheless, the transition from one pattern to another is so fast that it
possible get into a stationary limit cycle for the modulo-$q$ checking of $%
\mu $ in $\xi _i^\mu $.

\section{\protect\large The performance parameters}

For measuring the retrieving ability of this model, in a similar way to the
definition in Hopfield models, the overlaps of state $\mathbf{s}\left(
t\right) $ and the stored patterns with the initial arbitrary pattern $%
\mathbf{s}\left( 0\right) \in \left\{ -1,1\right\} ^N$is defined by 
\begin{equation}
m^\mu \left( t\right) =\frac 1N\sum_{j=1}^N\xi _j^\mu s_j\left( t\right)
,\qquad \mu =1,\ldots ,q
\end{equation}
For example, in $100$ neurons with $10$ stored patterns in a cycle, we
obtained a pattern sequence as $8\rightarrow 7\rightarrow 6\rightarrow
5\rightarrow 4\rightarrow 3\rightarrow 2\rightarrow 1\rightarrow
10\rightarrow 9\rightarrow 8\rightarrow 7\cdots $ (see Fig. 1a). It is easy
to find that the series is equal to a cycle of stored patterns with 10 
patterns for different retrieving times.

There are another two important parameters for describing long time
behaviors of networks: the cycle size $p$ and the relaxation time $r$. They
are defined by$^{9-11}$

\begin{equation}
p=\min_n\left[ \mathbf{s}\left( t+n\right) =\mathbf{s}\left( t\right) \right]
\end{equation}
\begin{equation}
r=\min_n\left[ \mathbf{s}\left( p+n\right) =\mathbf{s}\left( n\right) \right]
\end{equation}
Here, $p$ is the periodicity of attractors and $r$ is the time the system
converging into the $p$-period cycles.

In fact, the values of $p$ and $r$ are dependent upon the initial patterns $%
\mathbf{s}\left( 0\right) $ and the coupling matrix $\mathbf{J}$\textbf{\ }%
or the stored patterns $\mathbf{\xi }^\mu $. So, for a given value $p$, we
can choice many connection matrices\textbf{\ }through randomly generating a
large number of stored patterns. Moreover, the average size of cycles and
the average relaxation time can be defined by$^{9,10}$ 
\begin{equation}
\left\langle p\right\rangle =\frac 1M\sum_{n=1}^Mp\left( n\right)
\end{equation}
\begin{equation}
\left\langle r\right\rangle =\frac 1M\sum_{n=1}^Mr\left( n\right)
\end{equation}
in which $p\left( n\right) $ and $r\left( n\right) $ are defined by $(5)$
and $(6)$ for the $n$-th sample.

Now, viewing the stored patterns series as a whole, our interest is the
overlaps between $\mathbf{s}\left( t\right) $ and the stored cycle, not with
a simple pattern in the stored series. As a special case, only for $p=q$, we
introduced the average cycle overlaps 
\begin{equation}
\left\langle m\right\rangle \left( T\right) =\frac 1q\sum_{i=1}^q\max m^\mu
\left( Tq-i+1\right) ,\qquad \mu =1,\ldots ,q,\quad T=1,2,\ldots
\end{equation}
in which $T$ is the refreshed time scale. Figure 1b shows that the system
plotted in Figure 1a is successful in forming $\left\langle m\right\rangle
=1 $ for $q=10$, but there is a very small error with $\left\langle
m\right\rangle =0.963$ for $q=20$.

\section{\protect\large Simulation results}

For searching the features of period and relaxation time of attractors, we
set the $c_{ij}=1$ corresponding to the standard sequence processing
networks in fig. 2 to 4. We focus attention on the relations $\left\langle
p\right\rangle \, vs.\, \alpha $ and $\left\langle r\right\rangle \, vs. \,
\alpha $ in which the loading ratio $\alpha $ is defined by $q=\alpha N$.
>From comparing the value of $M$, the numbers of samples, given by $1000$ or 
$200$, we find simulation results of $\left\langle p\right\rangle \, vs. \,
\alpha $ have only a very small error. Considering our computer device, the
value of $N$ ranges from $50$ to $150$ and the value of $M$ is $200$ in the
following calculation.

Obviously there is a turning point $\alpha _c$ dividing the curve of $%
\left\langle p\right\rangle \, vs. \, \alpha $ into two parts (see Fig. 2).
In the first part with $\alpha <\alpha _c$, we find stable behaviors of
cycles with $\left\langle p\right\rangle \approx \alpha N$ equal to $q,$ the
number of stored patterns. Beyond this point, the curve gradually changes
into an exponential dependence of $\left\langle p\right\rangle /\left(
\alpha N\right) \propto \exp \left( A\alpha \right) $. Here, $A$ is a
constant in the range from $8.26$ to $46.74$ for $N=50$ to $150$. The value
of $\left\langle p\right\rangle $ increases so drastically that no results
of $\alpha >0.35$ are plotted. Additionally, the $\alpha _c$ are in the
range 0.13 to 0.17, and respectively the formation ratio $k/M=0.80$ to $0.85$%
, where $k$ is the number of $q$-period attractors with $\left\langle
m\right\rangle \geq 0.90$ and $M=200$ is the number of samples (see Fig. 3).

Figure 4 shows the average relaxation times in fully coupled networks.
Similarly, an exponential dependence $\left\langle r\right\rangle \propto
\exp \left( B\alpha \right) $ was observed with $B = 12.01$, $18.98$ and $%
29.07$ for $N=50$, $80$ and $120$, respectively.

In order to investigate the robustness of networks within the limits of $%
\alpha _c$, simulation with diluted connection have been carried out through
changing the values of coupling degree $d$ or setting $c_{ij}$ belong to $%
\left\{ 0,1\right\} $ randomly. In Fig. 5, in the same way as figure 2, it
is easy to see that there exists a critical point $d_c=0.15,0.35$ and $0.60$
for $\alpha =0.05$, $0.10$ and $0.15$. For stronger dilution, 
the average periodicity grows towards a plateau higher than $\alpha N$.

In conclusion, for small systems with spin neurons, it can be seen that the
ratio of the average size of attractors $\left\langle p\right\rangle $ and
the numbers of stored patterns is $1$ as long as the loading ratio $\alpha
>\alpha _c$, and then it increases strongly following an exponential law $%
\exp \left( A\alpha \right) .$ The more the system size increases, the more
the ratio grows. From analyzing the average cycle overlaps $\left\langle
m\right\rangle $ (Fig. 3), there is a narrow quickly decreasing region near
the turning point of periodicity. The maximal stored capacity$^5$ at $\alpha
_s=0.23$ for $N=100$, corresponding to 
$k/M=0.66$, is much larger than at $\alpha_s=0.32$ ( when $k/M=0.005$ ),
under the condition that $\left\langle m\right\rangle \geq 0.90$. Moreover,
the average relaxation time depends exponentially 
on the loading ratio and it raises faster with larger systems.

Additionally, through the simulation of the coupling degree $d$ in networks,
it is clear that there is a narrow transition region between small and large 
$\left\langle p\right\rangle /(\alpha N)$ centered on an inflection point $%
d_c$. Moreover, the larger loading ratio corresponding to the larger value
of $d_c$ means its robustness of stored information is more unstable. Beyond
our expectation, one result is a stable large attractor for highly diluted 
systems, another is that possibly $d_c$ depends only on $\alpha $ for
several different system sizes. This needs more detailed work in the future.

Above all, it is easy to see, not as in the asymmetric model given by
Gutfreund et al., that the asymmetric factor is a major parameter$^{9-12}$,
that the loading ratio $\alpha $ is a very important factor in sequence
processing networks. In fact, this difference from the definition of 
asymmetric connection strength by Gutfreund et al.

\bigskip

Fig. 1. (a) pattern overlaps and updating time for $N=100$ with $10$ stored
patterns in a cycles. (b) the average cycle overlaps plotted against the
refreshed time scale $T$.

\bigskip

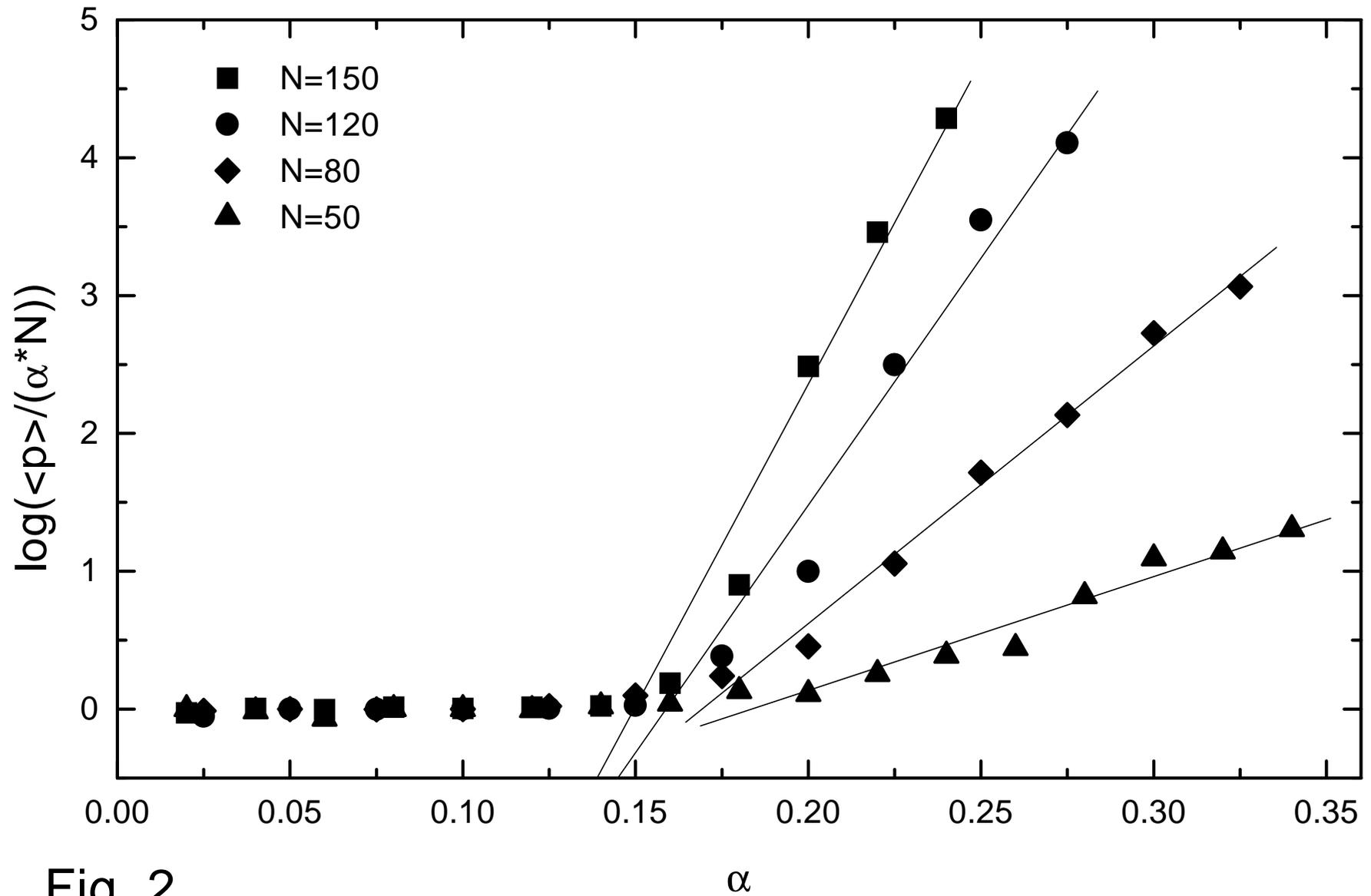
Fig. 2. logarithm of $\left\langle p\right\rangle /\left( \alpha N\right) $
against the loading ratio $\alpha $ for four different system size $N$.

\bigskip

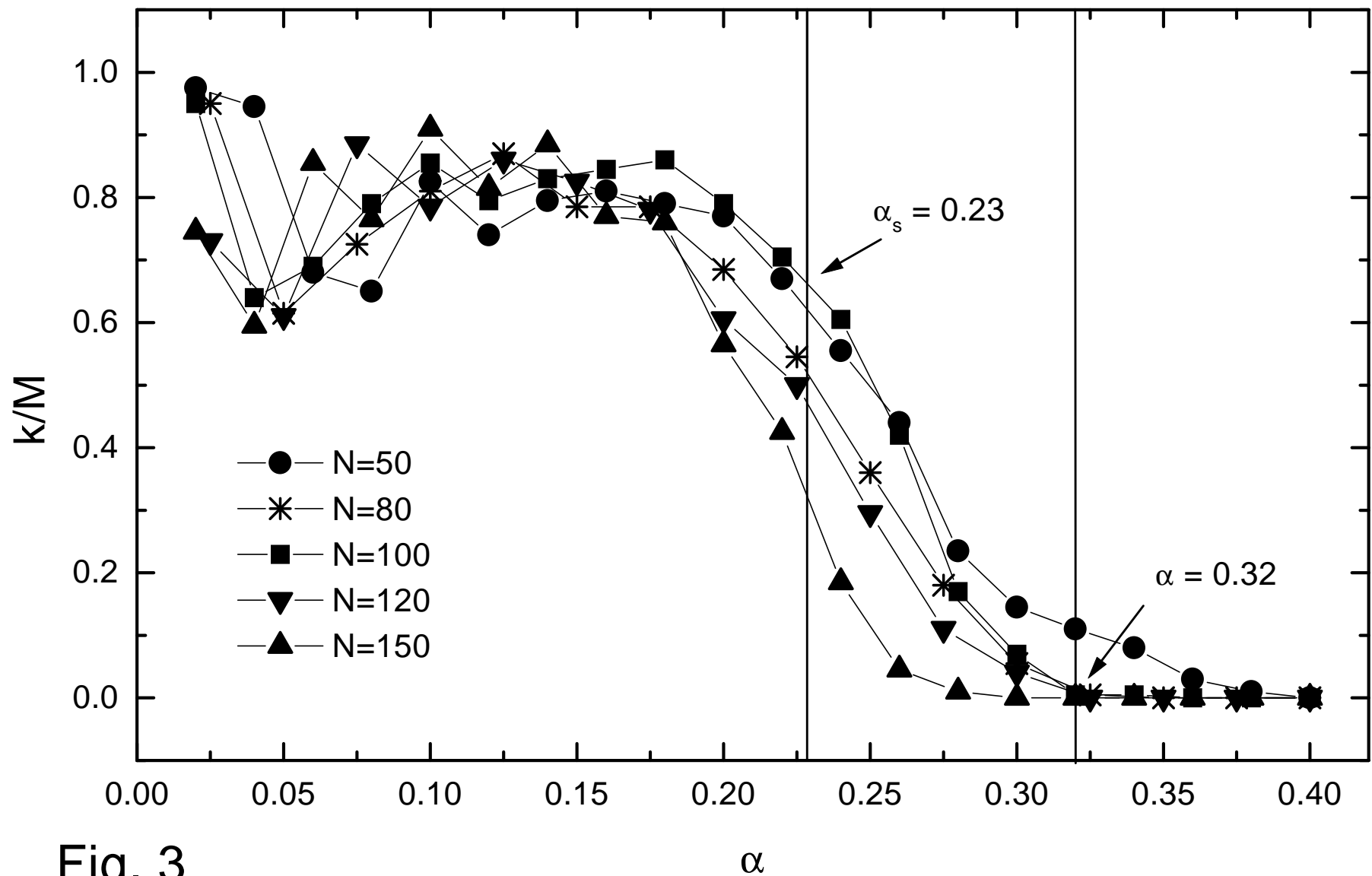
Fig. 3. the formation ratio $k/M$ for $\left\langle m\right\rangle \geq 0.90$
and $M=200$ against $\alpha $ for $\alpha <\alpha _c$ and $N=50$ to $150$.

\bigskip

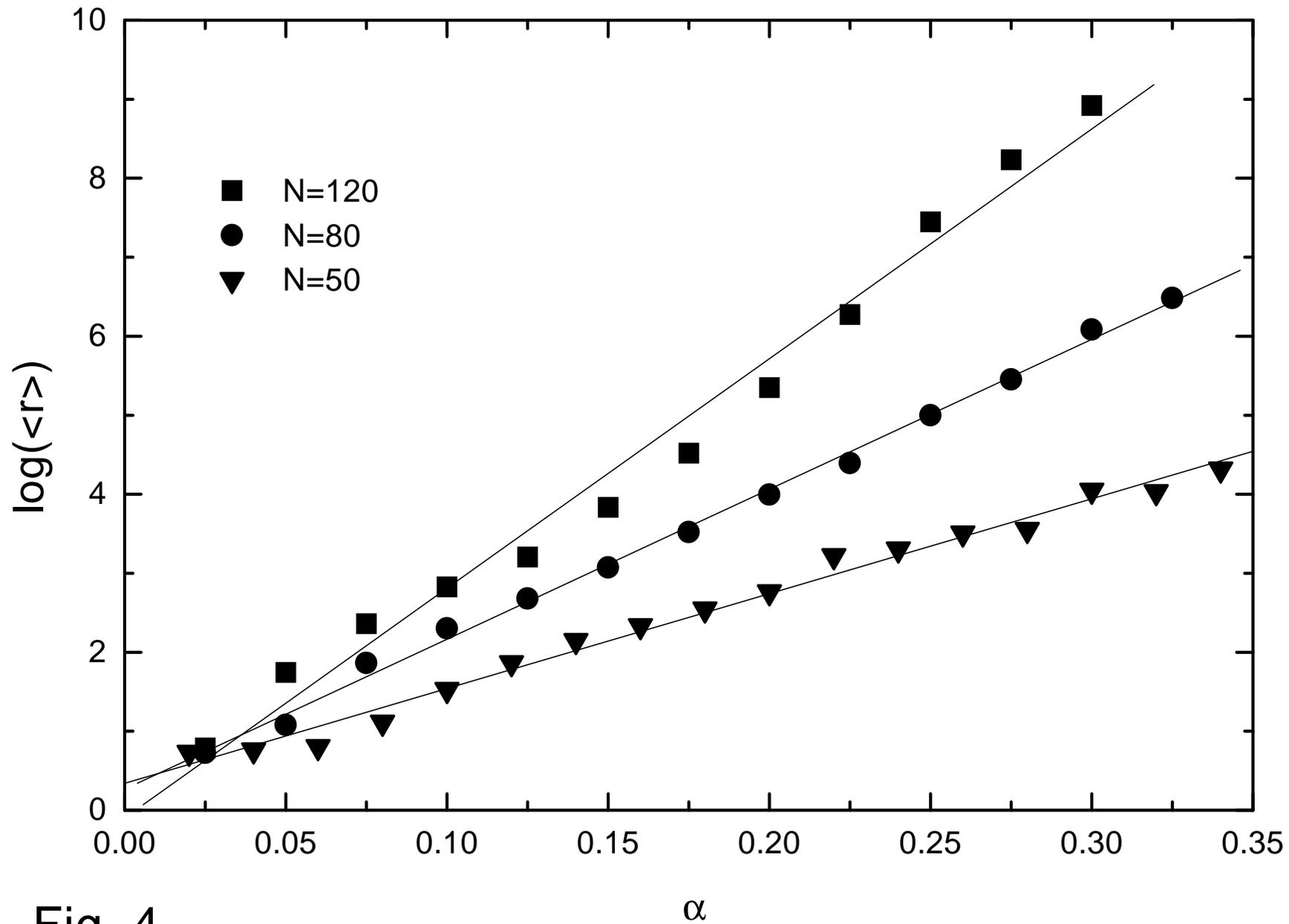
Fig. 4. logarithm of $\left\langle r\right\rangle $ against the loading
ratio $\alpha $ for three different system size.

\bigskip

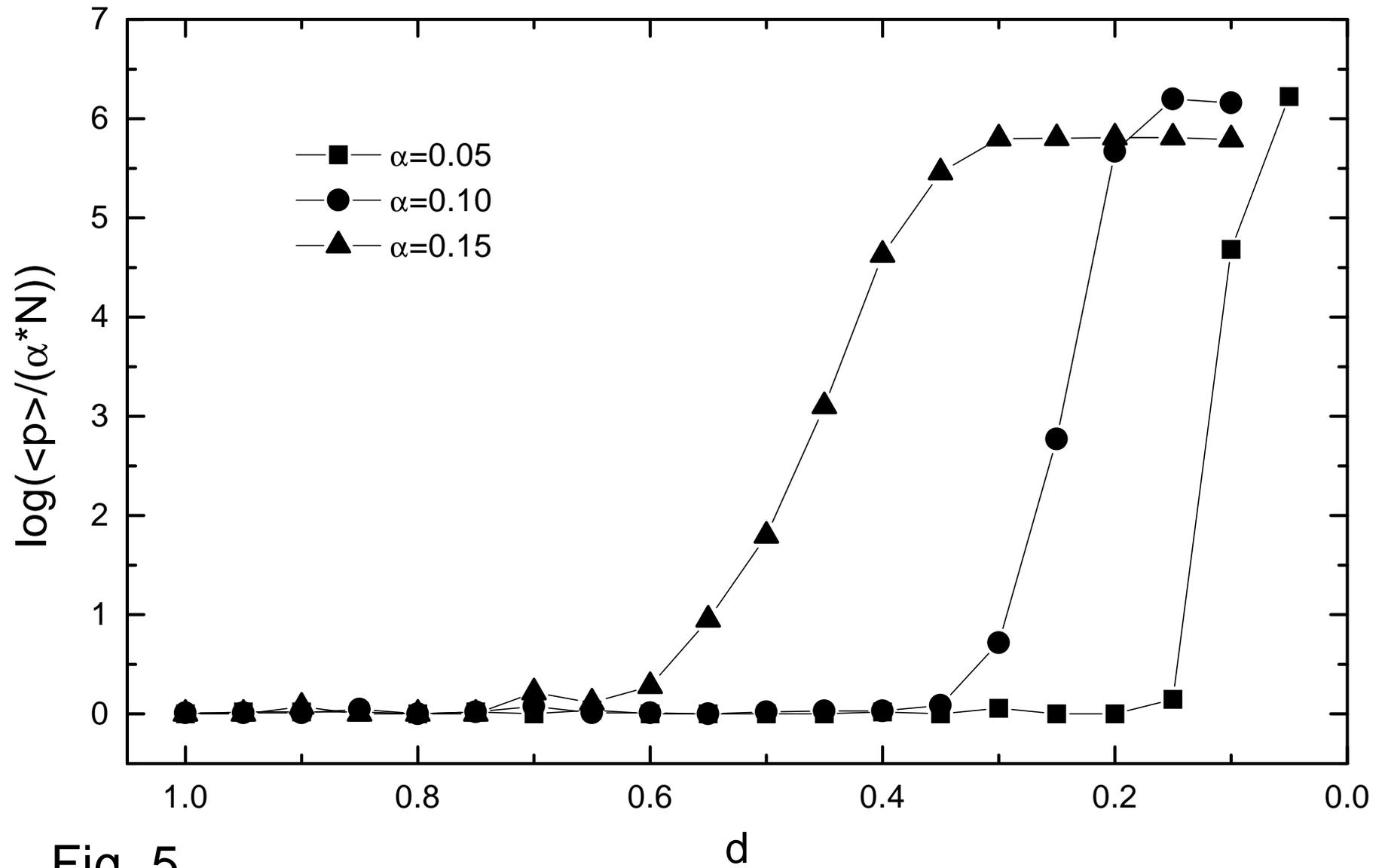
Fig.5. logarithm of $\left\langle p\right\rangle /\left( \alpha N\right) $
against the coupling degree $d$ for loading ratio $\alpha $ for $N=100$.

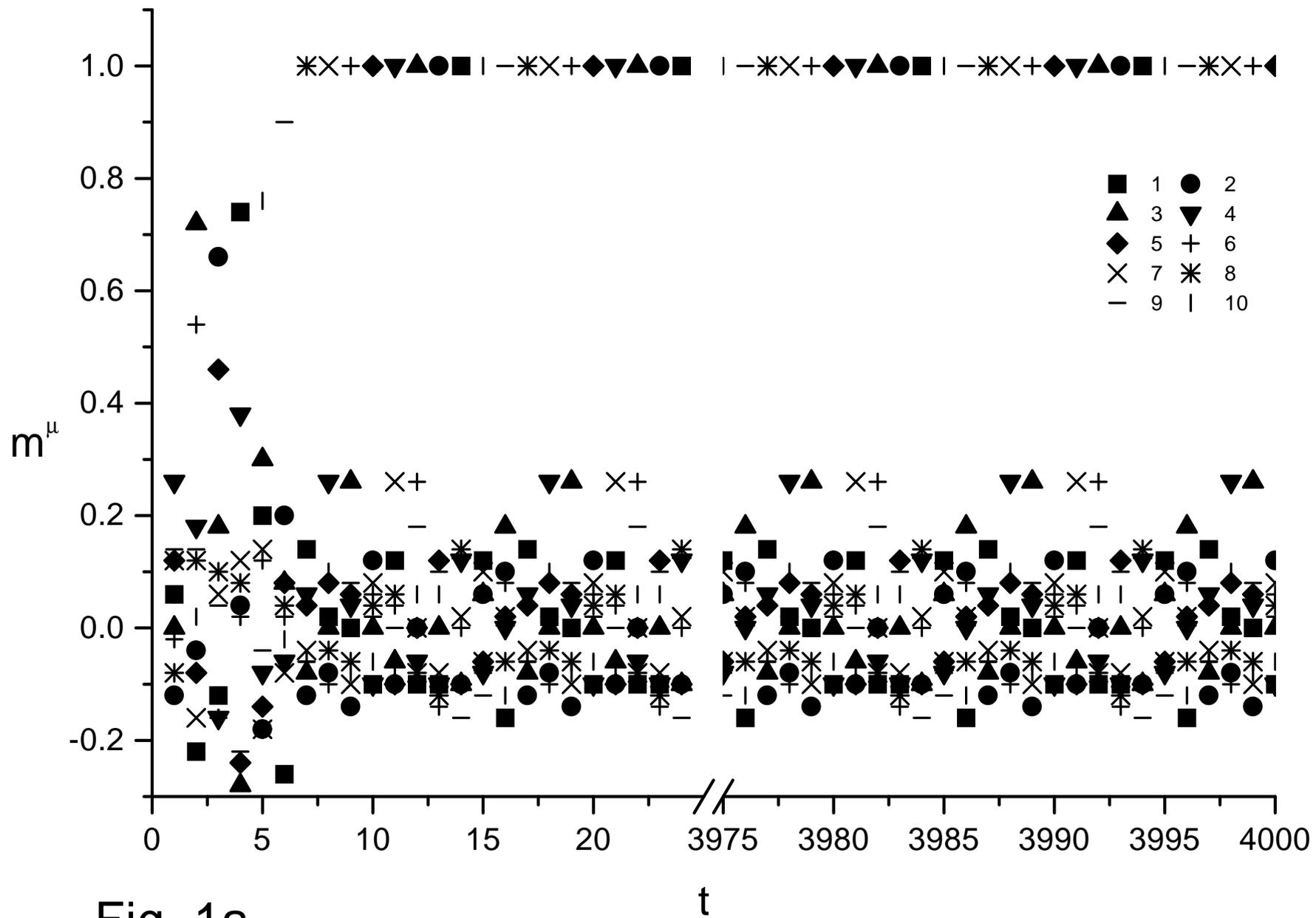
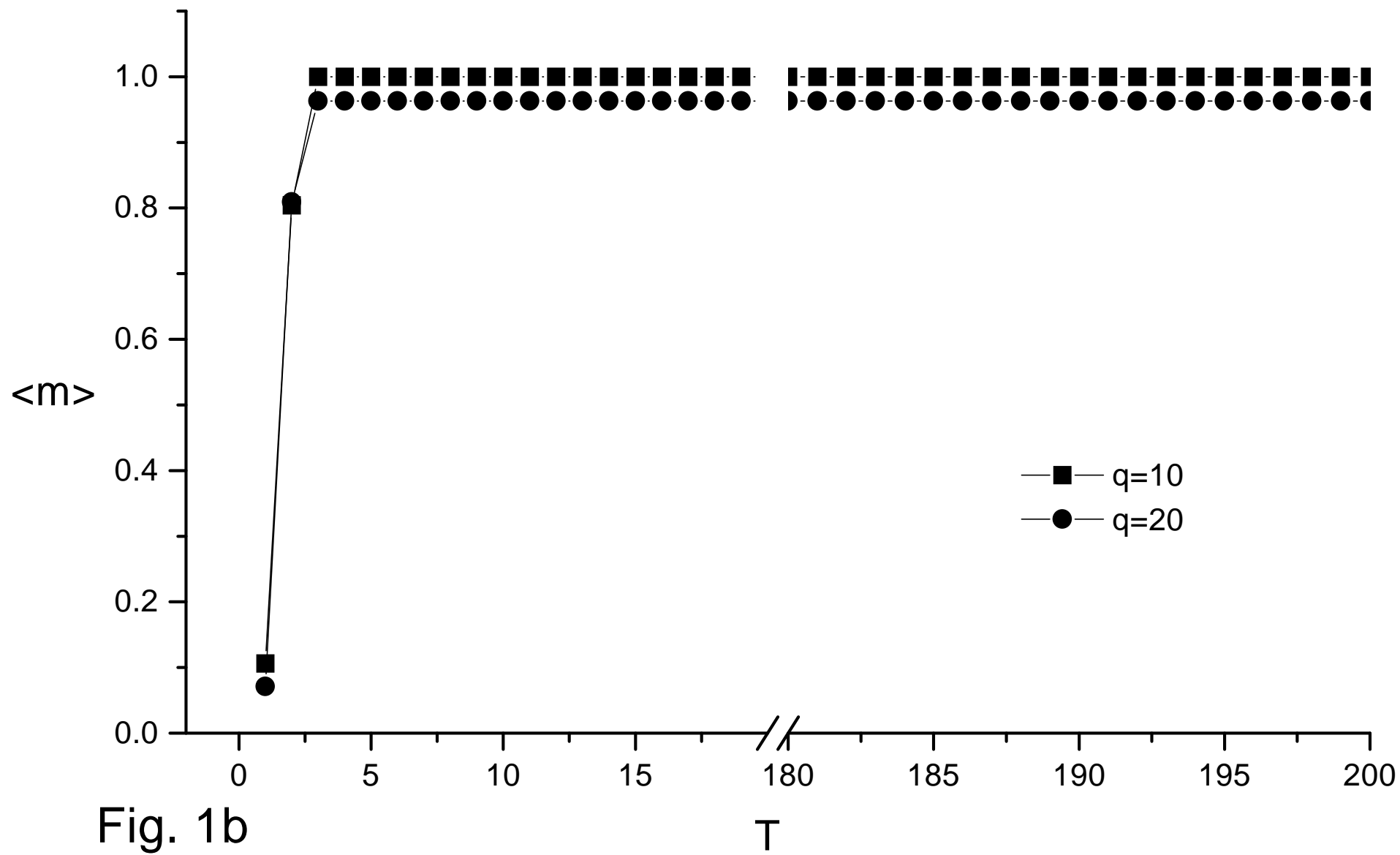
\end{document}